\documentclass[12pt]{article}
\usepackage{amsmath, amssymb,graphicx}

\newwrite\ffile\global\newcount\figno \global\figno=1

\def\writedef#1{}

\input epsf
\def\figin{\epsfcheck\figin}\def\figins{\epsfcheck\figins}
\def\epsfcheck{\ifx\epsfbox\UnDeFiNeD
\message{(NO epsf.tex, FIGURES WILL BE IGNORED)}
\gdef\figin##1{\vskip2in}\gdef\figins##1{\hskip.5in}
\else\message{(FIGURES WILL BE INCLUDED)}%
\gdef\figin##1{##1}\gdef\figins##1{##1}\fi}

\def\figinsert{}
\def\ifig#1#2#3{\xdef#1{fig.~\the\figno}
\writedef{#1\leftbracket fig.\noexpand~\the\figno}%
\figinsert\figin{\centerline{#3}}\medskip\centerline{\vbox{\baselineskip12pt
\advance\hsize by -1truein\center\footnotesize{  Fig.~\the\figno.} #2}}
\bigskip\endinsert\global\advance\figno by1}
\def\endinsert{}

\usepackage{amssymb}
\usepackage{graphics}
\begin{document}
\baselineskip 18pt
\newcommand{\Tr}{\mbox{Tr\,}}
\newcommand{\beq}{\begin{equation}}
\newcommand{\eeq}{\end{equation}}
\newcommand{\bea}{\begin{eqnarray}}
\newcommand{\eea}[1]{\label{#1}\end{eqnarray}}
\renewcommand{\Re}{\mbox{Re}\,}
\renewcommand{\Im}{\mbox{Im}\,}
\newcommand{\yms}{${YM^*\,}$}

\def\N{{\cal N}}


\thispagestyle{empty}
\renewcommand{\thefootnote}{\fnsymbol{footnote}}

{\hfill \parbox{4cm}{
        SHEP-05-15 \\
}}

\bigskip

\begin{center} \noindent \Large \bf
Towards a Perfect QCD Gravity Dual
\end{center}

\bigskip\bigskip\bigskip

\centerline{ \normalsize \bf Nick Evans, Jonathan P. Shock, Tom
Waterson \footnote[1]{\noindent \tt
 evans@phys.soton.ac.uk, jps@phys.soton.ac.uk, trw@phys.soton.ac.uk} }

\bigskip
\bigskip\bigskip

\centerline{ \it School of Physics and Astronomy} \centerline{ \it
Southampton University} \centerline{\it  Southampton, SO17 1BJ }
\centerline{ \it United Kingdom}
\bigskip

\bigskip\bigskip

\renewcommand{\thefootnote}{\arabic{footnote}}

\centerline{\bf \small Abstract} Many examples of gravitational
duals exist of theories that are highly supersymmetric and conformal
in the UV yet have the same massless states as ${\cal N}=2,1,0$ QCD.
We discuss such theories with an explicit UV cutoff and propose
that, by tuning higher dimension operators at the cutoff by hand,
the effects of the extra matter states in the UV may be removed from
the IR physics. We explicitly work in the AdS-Schwarzschild
description of QCD$_4$ and tune the operator Tr$F^4$ by relaxing the
near horizon limit to reproduce the lattice $0^{++}$ glueball mass
results. We find that to reproduce the lattice data, the IR and UV
cutoffs lie close to each other and there is essentially no
AdS-like period between them. The improved geometry gives a better
match to the lattice data for $0^{-+}$ glueball masses.

\newpage


\section{Introduction}

The AdS/CFT Correspondence \cite{hep-th/9711200,hep-th/9802109,hep-th/9802150} provides a dual gravitational
description of large $N$ SU($N$), ${\cal N}=4$ super Yang-Mills
theory. The extra fifth non-compact direction of the AdS space
corresponds to the energy scale of the gauge theory. Since the
gauge theory is conformal, and if it is at strong coupling, the
gravity description exists on an infinite line in this fifth
direction. Considerable work \cite{hep-th/9903026,hep-th/9909047,hep-th/0004063,hep-th/0003136} has been performed on studying the
inclusion of relevant operators in the field theory which
correspond to fields in AdS whose solutions fall to zero at large
radius in AdS. In this way one can study theories that have the
massless states of ${\cal N}=2,1,0$ theories. At large radius the
space returns to AdS and the gauge theory to the ${\cal N}=4$
theory. Typically in the interior the supergravity description
becomes singular providing an infra-red block which corresponds to
the induced mass gap of the theory with less supersymmetry.

Such constructions have been used as tools to study the low energy
behaviour of the ${\cal N}=2,1,0$ gauge theories and are often used
for comparison to the pure versions of those theories without the
extra states in the UV. Confinement
\cite{hep-th/9803002,hep-th/9803135,hep-th/9803001} and chiral symmetry breaking
\cite{hep-th/0306018,hep-th/0311270} are broadly well described. In
fact the bound state spectrums of these theories \cite{hep-th/9806021,hep-th/9810186,hep-th/9811156,hep-th/9806125,hep-th/0003115,hep-th/0209080,hep-th/0212207} also seem to match
well to those observed (and predicted by the lattice) in real QCD.
Recently a number of authors have constructed toy phenomenological
models of QCD using a slice of AdS space with appropriate fields to
describe the low lying hadronic states
\cite{hep-ph/0501128,hep-ph/0501218,hep-th/0501022}. These models
find good agreement with QCD at the 20-30$\%$ level or better.

Whilst deformed AdS geometries presumably do a good job of
catching QCD-like physics in the IR below the mass of the superpartners, these theories all have additional massive states at
strong coupling and evolve to a conformal strongly coupled theory
in the UV. A priori this appears to leave very non QCD-like
theories and any match with QCD states would appear to be telling
us mainly about the universality of these masses across a range of
gauge theories. In this paper we want to begin addressing the
issue of systematically removing this unwanted UV physics. We
clearly do not want a large UV strongly coupled conformal interval
so we will apply a hard UV cutoff in the gravitational
description corresponding to roughly the scale where QCD would
transition between perturbative and non-perturbative physics. In
the theories developed to date there will be additional fields to
those we want even at this scale (typically with masses of order
this scale). The couplings of these fields will necessarily alter
the physics of the fields we are interested in describing.  Since
the extra fields are massive we hope that their influence on the
running of the gauge coupling will be small. Their main effect
will be to distort the coefficients of higher dimension operators
in the fields we wish to study. We want to assume that the physics
above the cutoff scale is that of QCD rather than the ${\cal
N}=4$ theory but the higher dimensional operators will be the
wrong ones for this case if we just impose a cutoff. The natural
correction is to hand tune the higher dimension couplings to the
values in QCD to reproduce the correct physics. This is what we
begin to study in this letter.

The idea of tuning higher dimension operators to remove the
effects of ``regulator'' fields is similar to the idea of {\it
perfect} or {\it improved} actions in lattice gauge theory \cite{hep-lat/9308004,Luscher:1984zf}. If
working on too coarse a lattice, the lattice infects the QCD
physics under study with artifacts. However, in principle, by
appropriately tuning the higher dimension operators of the theory
it should be possible to precisely reproduce QCD results even on
an arbitrarily coarse lattice. In practice, adjusting just one or
a few higher dimension operators to correctly reproduce the
physical data shows improvement across the whole predicted
spectrum.

There are also strong links to the ideas of the exact
renormalization group - a number of studies have been made
\cite{hep-th/0006064,hep-th/0106258} where a gauge theory of
interest is UV regulated by either ${\cal N}=4$ Yang Mills'
conformal nature or by Pauli Villar's type fields. If the mass of
these fields is at a scale where there is weak coupling they
essentially decouple. However, the exact renormalization group
provides a formalism that tracks how one must switch on higher
dimension operators to keep the physics invariant even as these
fields are made light and brought into the energy regime of strong
coupling.

In this paper we will study the glueball mass spectrum of the
AdS$_7$ Schwarzschild black hole. This solution describes the theory
on the world volume of an M5 brane with a compact dimension and at
finite temperature. The resulting theory is believed, below the
temperature scale, to describe four dimensional non-supersymmetric
Yang Mills theory. The glueball spectrum is well known
\cite{hep-th/9806021,hep-th/9810186,hep-th/9811156,hep-th/9806125,hep-th/0003115,hep-th/0209080,hep-th/0212207}
and lies within 30$\%$ or so of the QCD lattice results. The
additional M5 brane fields have masses of order the temperature of
the field theory and distort the UV of the theory into a strongly
coupled 5+1 dimensional theory. An attempt was made in
\cite{hep-th/9810186} to remove some of the extra fields in the
theory by making the brane configuration rotate. However, to
maintain a gravitational description, extra fields are needed to
keep the theory at strong coupling in the UV so they can not be
removed completely.  Instead we will introduce an explicit UV cutoff into this theory and then tune the coefficient of the Tr$F^4$
coupling at the cutoff to try to remove the incorrect UV physics
(i.e. the effects of these extra fields) in the glue sector and
hence improve the glueball results. To switch on this operator one
simply allows the solution to revert to flat space asymptotically by
undoing the near horizon limit. The resulting deformation has the
correct dimension and symmetry properties to play the role of the
Tr$F^4$ coupling. Previous studies of this theory can be found in
\cite{hep-th/0112058,hep-th/9909082, hep-th/9903227,hep-th/9905081}; the operator may result from any number of
changes to the UV of the theory, from switching on gravity to
embedding the theory in a multi-centre solution. We imagine that
above our UV cutoff the theory is true Yang Mills theory. If we did
not impose a cutoff the higher dimension operator would grow into
the UV and eventually come to dominate the physics. In this case the
operator makes the potential that is responsible for the discrete
glueball spectrum unbounded.  We therefore take the scale where the
potential instability sets in as the natural UV cutoff.

We will see that to match the lattice large $N$ $0^{++}$
glueball\footnote{Throughout the text we will label the glueballs by
their quantum numbers $J^{PC}$} mass data, we must make the operator
$\Tr F^4$ large at a rather low scale. In fact, there turns out to
be such a small interval between the UV and IR cutoffs that there
is no AdS like geometry left and barely any gravity description at
all! This is perhaps not surprising since QCD presumably moves into
the strong coupling regime fairly quickly and then almost
immediately generates a mass gap. We nevertheless look at the
predictions of our short interval for the $0^{-+}$ glueballs. Only
$N=3$ lattice data exists but our improved geometry is a better
match to the data than the unimproved geometry.

\section{The Improved Geometry}
To construct a dual gravitational description of QCD$_4$ we
begin with the M5 brane solution of 11d Euclidean supergravity \cite{Nucl.Phys.B360.197}. 
\beq \label{eq:11dNE}
ds_{11}^2=h^{-\frac{1}{3}}\left[\left(1-\frac{b^6}{\rho^3}\right)d\tau^2+\sum_{i=1}^5dx_i^2\right]+h^\frac{2}{3}\left[\left(1-\frac{b^6}{\rho^3}\right)^{-1}d\rho^2+\rho^2d\Omega_4^2\right],
\eeq
where $h$ is the solution of the 5d Laplace equation and $b$ corresponds to the inverse of the temperature of the dual field theory \cite{hep-th/9803131}. If we let $\rho=\lambda^2$ and go to the non-extremal near horizon limit ($b=0$, $h=\rho^{-3}$) of this metric, we get:
\beq
ds_{11}^2=\lambda^2\left[d\tau^2+\sum_{i=1}^5dx_i^2\right]+\frac{4}{\lambda^2}d\lambda^2+d\Omega_4^2,
\eeq
ie. AdS$_7\times${\bf S}$^4$ after appropriate scaling of coordinates and a Wick rotation. From this we can see that $\rho$ has mass dimension two. 

We now want to modify the metric to include the effect of adding $\Tr F^4$ to the
dual field theory. \beq
  S_{FT}=\int d^6 x \left[\frac{1}{g^2}\Tr F^2 + G \Tr F^4 +
  \cdots\right]
\eeq The coupling $G$ has mass dimension -6. The gravitational dual
of this can be included by adding a constant term in the solution
for $h$
\beq
  h=\rho^{-3}\rightarrow\rho^{-3}(1+\alpha\rho^3),
\eeq i.e. going away from the near horizon limit. $\alpha$ has the
right mass dimension -6 to be dual to $G$ plus correctly has no
R-charge since it does not depend on angles on the four sphere.

QCD$_4$ is dual to the low energy limit of IIA string theory on the AdS-Schwarzschild background \cite{hep-th/9905111, hep-th/9802042} on imposing anti-periodic boundary conditions for the fermions in the compact $\tau$ direction. We can obtain the type IIA metric from (\ref{eq:11dNE}) by compactifying the $11^{th}$ dimension and rescaling the metric by a factor $e^\frac{2\phi}{3}=h^{-\frac{1}{6}}$:
\beq \label{eq:stringmetric2}
ds_{\rm IIA}^2=h^{-\frac{1}{2}}\left[\left(1-\frac{b^6}{\rho^3}\right)d\tau^2+\sum_{i=1}^4dx_i^2\right]+h^{\frac{1}{2}}\left[\left(1-\frac{b^6}{\rho^3}\right)^{-1}d\rho^2+\rho^2d\Omega_4^2\right].
\eeq
This solution has a non-constant dilaton $e^{\phi}=h^{-\frac{1}{4}}$.
This is our improved geometry in which we will now calculate the
glueball spectra.

Note that the function $h$, being a solution of the five dimensional
Laplace equation, can encode a more complicated function if we
allow it to have angular dependence. Terms in $h$ that fall off at
large radius are associated with operators of the form $\Tr \phi^n$
in the field theory \cite{hep-th/9811120, hep-th/0105235}, whilst those that grow correspond to R-charged
higher dimension operator couplings. Since we are interested in the
glue sector we will not make use of these operators.

\section{The $0^{++}$ Mass Spectrum}
We first calculate the mass spectrum of the $0^{++}$ glueball. This
is derived on the supergravity side by solving the equation of
motion for a massless scalar
\beq
  \partial_{\mu}\left[\sqrt{g}g^{\mu\nu}e^{-2\phi}\partial_{\nu}\Phi\right]=0
\eeq in the string frame background (\ref{eq:stringmetric2}). Assuming that
$\Phi$ is of the form $\Phi = f(\rho)e^{i k \cdot x}$ this gives
\beq\label{eq:ppwave}
  \frac{1}{\rho}\frac{d}{d \rho}\left[(\rho^4-\rho)\frac{d f}{d \rho}\right]= k^2(1+\alpha\rho^3)f(\rho),
  \eeq
where we have set $b=1$ and the mass of the dilaton is $m^2=-k^2$ in
units of $b$.

In order to change this equation of motion into a Schr\"odinger
form, we make a change of the dependant variable to $z$ and
rescale $f$ \beq
  \frac{d z}{d \rho}=\sqrt{\frac{1+\alpha\rho^3}{\rho^3-1}}, \hspace{1cm}
  f(z)\rightarrow f(z)e^{-\frac{1}{2}\int d z' p(z')}, \eeq where
\beq
  p(\rho)=\frac{5\rho^3-2+\alpha\rho^3(8\rho^3-5)}{2\rho\sqrt{\rho^3-1}(1+\alpha\rho^3)^{\frac{3}{2}}}.
\eeq We now have an equation in Schr\"odinger form with a
potential

\beq
  -g''(z)+Q(z)g(z)=m^2g(z), \hspace{1cm}
 Q(z)=\frac{1}{2}p'(z)+\frac{1}{4}p(z)^2,
\eeq
or, in terms of the radial coordinate $\rho$
\beq
  Q(\rho)=\frac{1}{2}\sqrt{\frac{\rho^3-1}{1+\alpha\rho^3}}\frac{d p}{d\rho}+\frac{1}{4}p(\rho)^2.
\eeq We plot the potential as a function of $\alpha$ in Figure 1 for
different values of $\alpha$. For $\alpha=0$ the pure AdS geometry
gives a well that is bounded into the UV and an infinite, discrete
glueball spectrum. When $\alpha$ is non-zero, the UV potential is
modified and eventually falls to zero. In the field theory the
higher dimension operator grows into the UV until it dominates the
physics and removes the discrete spectrum. If we allow this to
happen then we are not describing a QCD-like theory in the UV, so
we will impose a hard UV cutoff, $\Lambda$. The natural scale to
place the cutoff is at the turning point of the potential since
that includes in the IR theory the highest possible tower of
discrete states - we will adopt this value for the cutoff
henceforth. Thus as we increase $\alpha$ we will necessarily be
working on a shorter radial interval.

\begin{figure}[!h]
\begin{center}\label{f:potential}
\includegraphics[height=7cm,clip=true,keepaspectratio=true]{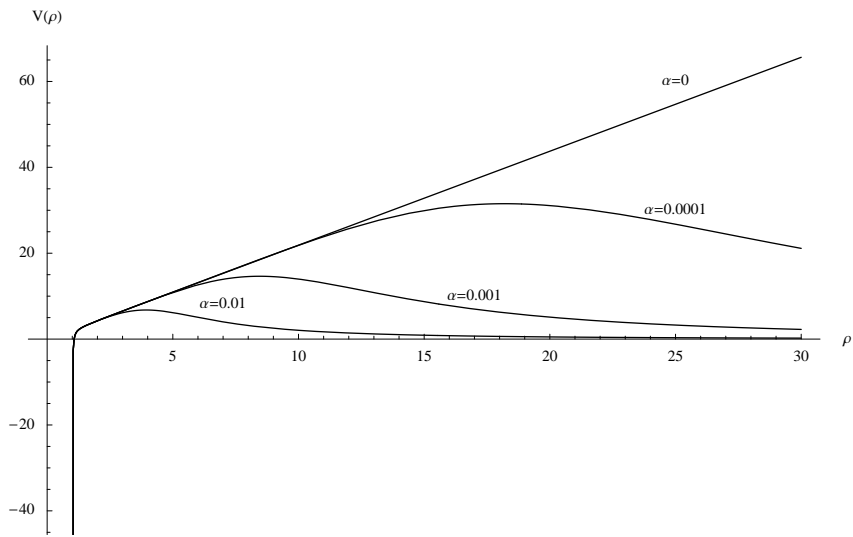}
\caption{QCD$_4$ Schr\"odinger potential for the $0^{++}$ glueballs $\alpha=0$ to
$\alpha=0.01$.}
\end{center}
\end{figure}

We calculate the eigenvalues of (\ref{eq:ppwave}) numerically
using the ``shooting technique'', whereby we fix a value for
$k^2$, set boundary values for $f'(\rho)$ at the cutoff $\Lambda$
and then solve the equation of motion numerically down to the
black hole horizon at $\rho=1$. We then repeat this for different
values of $k^2$. The eigenvalues will be those values of $k^2$ for
which the solution is regular at the horizon i.e. for which $f'(0)$
is constant.

We know that for $\alpha=0$ taking $\rho\rightarrow\infty$, the
metric goes to AdS space and the normalisable solution to the wave
equation (\ref{eq:ppwave}) goes like $\rho^{-3}$. We will use the
naive boundary conditions $f(\Lambda)=\Lambda^{-3}$,
$f'(\Lambda)=-3\Lambda^{-4}$ in all cases below. As $\alpha$ grows
and the cutoff falls this boundary condition, which represents
the effective dimension of Tr$F^2$, should presumably change  - it
is essentially a matching condition on the dimension that should
come from the UV theory. We have checked that if we instead use
the boundary conditions $f(\Lambda)=\Lambda^{-(3+\epsilon)}$,
$f'(\Lambda)=-(3+\epsilon)\Lambda^{-(4+\epsilon)}$ with $-1 <
\epsilon< 1$ the ratio of the lightest two glueballs masses only
changes by 6\%. This indicates that the mass spectrum is largely
insensitive to the precise values of the boundary conditions. Note
that this range includes  Tr$F^2$ having dimension four as one
might expect in real QCD.

Using this shooting technique, we tune $\alpha$ to get a glueball
spectrum that agrees best with the available large $N$ lattice
data \cite{hep-lat/9901004,hep-lat/0103027}. Figure 2 shows the ratio
$m(0^{++*})/m(0^{++})$ for different values of $\alpha$. We can
see that setting $\alpha=0.0855$ gives the correct value for the
second glueball mass (the first is fixed by normalisation). This
implies that $\Lambda=1.99$ - we will refer to this case as the
``improved geometry". For this value of $\alpha$ we get the
spectrum of masses shown in Table 1. The glueball masses rise in
the theory although we have no more lattice data to compare to for
this state.

The result, that to correctly reproduce the lattice data we must
raise $\alpha$ so that the theory only provides a description
between an IR scale of $b=1$ and a UV scale of $\sqrt{\Lambda}=1.41$ (note $\sqrt{\Lambda}$ has mass dimension 1), is
important. Although the original AdS black hole produced results
that match the QCD data reasonably we find that to move to a
phenomenological model of QCD we must actually distort the AdS space
considerably. Indeed, the gravitational theory's interval is
worryingly small and non-AdS like. This is not so surprising in
terms of real QCD where the regime between the QCD coupling becoming
non-perturbative and the scale of the mass gap of the theory is
quite small. This result may have important ramifications for
attempts to turn toy models of the sort in
\cite{hep-ph/0501128,hep-ph/0501218,hep-th/0501022} into true
phenomenological tools.

\begin{center}
\begin{tabular}{||l|l|l|l|l||}\hline
\emph{Glueball State} & \emph{Improved Geometry} &
\emph{$\alpha=0$} &\emph{$N=3$ Lattice} & \emph{$N=\infty$
Lattice}\\ \hline $0^{++}$ & 1.00 & 1.00 & 1.00 & 1.00\\ \hline
$0^{++*}$ & 1.90 &
1.58 & 1.74 & 1.90\\ \hline $0^{++**}$ & 3.05 & 2.15 & - & -\\
\hline $0^{++***}$
& 4.27 & 2.72 & - & -\\ \hline $0^{++****}$ & 5.52 & 3.33 & - & -\\
\hline
\end{tabular}

\vspace{0.2in}
Table 1: QCD$_4$ $0^{++}$ glueball masses from AdS ($\alpha=0$) and Improved ($\alpha=0.0855$) geometries along with lattice data \cite{hep-lat/9901004,hep-lat/0103027}. Normalisation is such that the ground state mass is set to one.
\label{tab:plusplus}
\end{center}

\begin{figure}[!h]
\begin{center}\label{f:ratiovsalpha}
\includegraphics[height=7cm,clip=true,keepaspectratio=true]{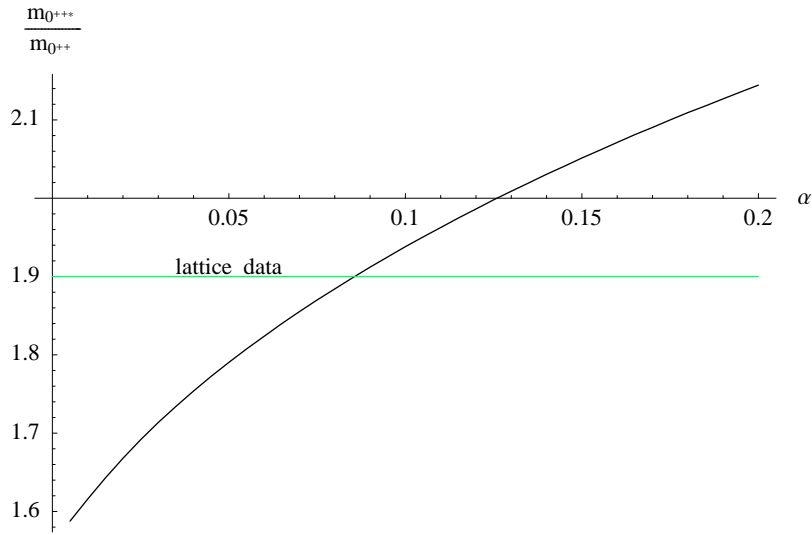}
\caption{$m_{0^{++*}}/m_{0^{++}}$ for different values of $\alpha$}
\end{center}
\end{figure}

\section{The $0^{-+}$ Mass Spectrum}

We now consider the $0^{-+}$ glueballs. We assume that the dominant
contribution to their mass spectrum will come from the $\Tr F\tilde
F$ operator as this is the lowest dimension operator with the
correct quantum numbers. Dual to this on the supergravity side is
the RR 1-form $A_{\mu}$. The equation of motion for this is
\beq
  \partial_\nu\left[\sqrt{g}g^{\mu\delta}g^{\nu\sigma}(\partial_\delta A_\sigma-\partial_\sigma A_\delta)\right]=0.
\eeq As with the $0^{++}$, we look for solutions of the form $A_\tau
= f(\rho)e^{i k \cdot x}$ in the background
(\ref{eq:stringmetric2}). The result is the equation
\beq
  \frac{1}{\rho^4}(\rho^3-1)\frac{d}{d \rho}\left[\rho^4\frac{d f}{d \rho}\right]=k^2(1+\alpha\rho^3)f(\rho).
 \eeq
If we set $\alpha=0.0855$ and $\Lambda=1.99$, which were their optimum
values for the $0^{++}$, we get the spectrum shown in Table 2. We
have normalised all masses to the $0^{++}$ ground state. The
lightest state does not match well to the N=3 lattice data - this
state was omitted from the spectrum in
\cite{hep-th/9810186,hep-th/9811156} which then improves the fit
considerably! The effect of our improved geometry is to make the
states more massive which improves the fit to the data whether the
first state is omitted or left in. To truly match these states to
the data would presumably require a higher dimension operator with
$P=-1,C=+1$ quantum numbers to be tuned - it is not clear how to
include such an operator though.

\begin{center}
\begin{tabular}{||l|l|l|l||}\hline
\emph{Glueball State} & \emph{Improved Geometry} &
\emph{$\alpha=0$} & \emph{$N=3$ Lattice} \\ \hline $0^{-+}$ &
0.35 & 0.29 & 1.61 \\ \hline $0^{-+*}$ & 1.38 & 1.24 & 2.26 \\
\hline $0^{-+**}$ & 2.48 & 1.84 & - \\ \hline $0^{-+***}$ & 3.71 &
2.42 & - \\ \hline
\end{tabular}
\vspace{0.2in}

Table 2: QCD$_4$ $0^{-+}$ glueball masses from AdS ($\alpha=0$) and Improved ($\alpha=0.0855$) geometries along with lattice data \cite{hep-lat/9901004} \label{tab:minusplus1}. All states are normalised to the $0^{++}$ ground state.
\end{center}

\section{Conclusions}
We have shown that by including the gravitational dual of a higher
dimensional field theory operator to the usual AdS-Schwarzschild
metric, we can tune our theory to match onto the first excited
$0^{++}$ glueball state as calculated using lattice techniques.
Having fixed the strength of this perturbation, we also find that
the $0^{-+}$ spectrum is improved. We find that in order to get the
correct $0^{++}$ spectrum we almost entirely remove the AdS-like
region of the space. This ties in with there being only a small
energy range between the mass gap and strong coupling region of QCD.

\hspace{0.5in}

\noindent {\bf Acknowledgements}

JS and TW are grateful to PPARC for the support of their
studentships. NE is grateful to the organisers of the ``AdS/CFT and QCD'' workshop at Hanyang university in Korea in October 2004 where the idea for this work was first developed.

\end{document}